\documentclass[12pt]{elsarticle}
\usepackage{amssymb,amsmath,graphicx}
\journal{Nuclear Physics B}
\allowdisplaybreaks
\begin{document}

\begin{frontmatter}




\title{Master Integrals for Four-Loop Massless Propagators up to Transcendentality Weight Twelve}
\author[binp]{R.N. Lee}
\ead{r.n.lee@inp.nsk.su}%
\author[srcc,kit]{A.V. Smirnov}
\ead{asmirnov80@gmail.com}
\author[sinp,kit]{V.A. Smirnov}
\ead{smirnov@theory.sinp.msu.ru}
\address[binp]{Budker Institute of Nuclear Physics and Novosibirsk State University, 630090, Novosibirsk, Russia}
\address[srcc]{Scientific Research Computing Center, Moscow State University, 119992 Moscow, Russia}
\address[sinp]{Skobeltsyn Institute of Nuclear Physics of Moscow State University, 119992 Moscow, Russia}
\address[kit]{Institut f\"{u}r Theoretische Teilchenphysik, KIT, 76128 Karlsruhe, Germany}

\begin{abstract}
We evaluate a
Laurent expansion in dimensional regularization parameter $\epsilon=(4-d)/2$ of
all the master integrals for four-loop massless propagators up to transcendentality weight twelve, using a recently developed method
of one of the present coauthors (R.L.) and extending thereby results
by Baikov and Chetyrkin obtained at transcendentality weight seven.
We observe only multiple zeta values in our results.
Therefore, we conclude that all the four-loop massless propagator integrals, with any
integer powers of numerators and propagators, have only multiple zeta values
in their epsilon expansions up to transcendentality weight twelve.
\end{abstract}

\begin{keyword}

multiloop Feynman integrals \sep dimensional regularization \sep
multiple zeta values
\end{keyword}

\end{frontmatter}
%
%
%\section{Introduction}

About one year ago Baikov and Chetyrkin published their results of evaluation of
all the master integrals for four-loop massless propagators in a Laurent expansion in
dimensional  regularization parameter $\epsilon=(4-d)/2$ up to transcendentality weight
seven~\cite{Baikov:2010hf}.

These integrals are associated with graphs depicted in Fig.~1.
The corresponding coefficients
at powers of $\epsilon$ turned out to be linear combinations of
multiple zeta values (MZV)
\begin{equation}\label{MZVdef}
\zeta(m_1,\dots,m_k)=\sum\limits_{i_1=1}^\infty\sum\limits_{i_2=1}^{i_1-1}
\dots\sum\limits_{i_k=1}^{i_{k-1}-1}\prod\limits_{j=1}^k\frac{\mbox{sgn}(m_j)^{i_j}}{i_j^{|m_j|}}
\end{equation}
(see, e.g., Ref. \cite{BBV}).
In our recent publication~\cite{Lee:2011jf}, we studied three four-loop non-planar massless
propagator integrals (corresponding to non-planar graphs) where, according to the
analysis of Brown~\cite{Brown:2008um}, one
could meet not only MZV but also Goncharov's polylogarithms \cite{Goncharov} with
sixth roots of unity as arguments. One of them is $M_{4,5}$ in Fig.~1 and the other two
integrals are not master integrals and reduce to some master integrals in this figure.
We performed calculations up to transcendentality weight twelve~\cite{Lee:2011jf} and
observed only MZV in results.

In this paper we continue our experimental investigation of
the four-loop massless propagator diagrams and present results for {\em all}
the master integrals up to transcendentality weight twelve.
Our motivation is twofold. First, we would like to check explicitly whether there are only MZV in results. Second, we would like to demonstrate further the power of the "dimensional-recurrence-and-analyticity" (DRA) method~\cite{Lee:2009dh} that we use. The method is  based on the use of dimensional recurrence relations (DRR) \cite{Tarasov1996} and analytic properties of Feynman integrals as functions of the parameter of dimensional regularization, $d$, and was already successfully applied in previous calculations
\cite{Lee:2010cga,Lee:2010ug,Lee:2010wea,Lee:2010hs,Lee:2010ik,Lee:2011jf}.

A necessary condition of the application of DRA method is the possibility to make an integra\-ti\-on-by-parts (IBP) \cite{IBP} reduction of integrals participating in DRR to master integrals.
To do this, we use the {\tt C++} version of the code {\tt FIRE} \cite{FIRE}. To reveal analytic properties of the master integrals
we used a sector decomposition \cite{BH,BognerWeinzierl,FIESTA}
implemented in the code {\tt FIESTA} \cite{FIESTA,FIESTA2}.
To fix remaining constants in the homogenous solutions of dimensional
recurrence relations it was generally not sufficient (contrary to our previous
work~\cite{Lee:2011jf} on the evaluation of three non-planar diagrams)
to use analytic
results for the four-loop massless propagators master integrals \cite{Baikov:2010hf}
(confirmed numerically by {\tt FIESTA} in Ref.~\cite{Smirnov:2010hd} where one more
term of the $\epsilon$-expansion was obtained)
up to transcendentality weight seven.
To obtain additional information here
we applied the method of Mellin--Barnes (MB) representation \cite{MB1,MB2,books2}. For each integral the number of terms calculated with the help of MB representation was sufficiently large to provide at least one consistency check of our results.

At the final stage of the method, we applied the PSLQ algorithm \cite{PSLQ} as well as the code {\tt HPL} \cite{Maitre:2005uu} for dealing with MZV. Since the rational coefficients at transcendental numbers turn out to be quite cumbersome we applied PSLQ with the rather high accuracy of 1500 digits.

\begin{figure}
  \centering
  \includegraphics[width=0.95\textwidth]{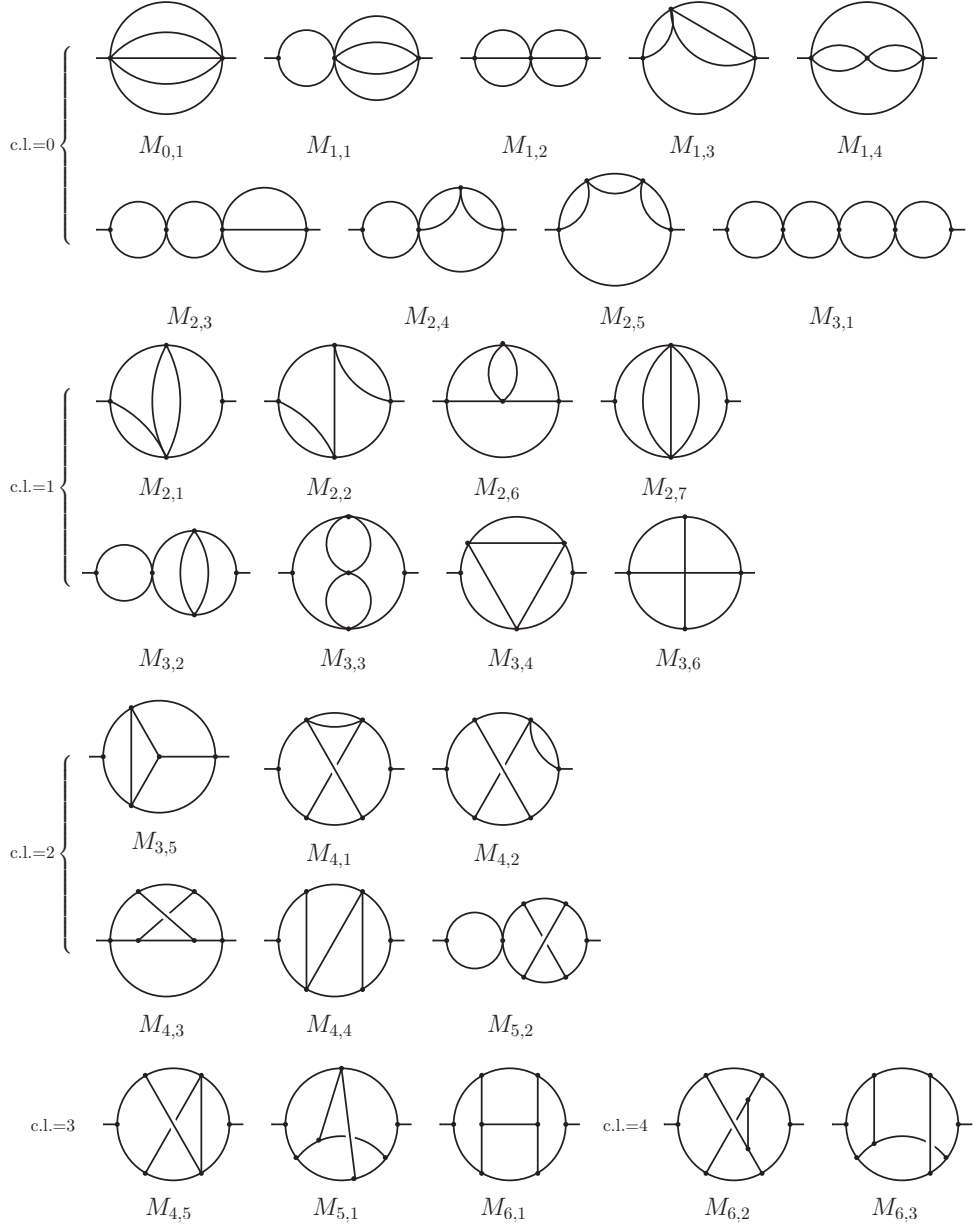}\\
  \caption{%Diagrams considered in this paper.
Master diagrams for four-loop massless propagators.
Here c.l. means the complexity level --- see
Ref. \cite{Lee:2010cga}.}\label{Fig:integrals}
\end{figure}
In our results presented below, we tried to reveal as much as possible the homogeneous transcendentality. This is achieved by pulling out suitable rational functions of $\epsilon$ and by considering a linear combination, with rational coefficients, of a given master integral and some of its lower master integrals. However, for the last three integrals, the task of finding such linear combination is quite
complicated because of the large number of the lower master integrals. For these three integrals we have succeeded to find their representation in the form of a sum of several homogeneous terms with rational coefficients in $d$. This representation allowed us to pull out a rational factor so that any transcendental number appeared only in the limited number of consecutive terms of $\epsilon$ expansion.
We choose the following loop integration measure
\begin{align}
\frac{\Gamma (d-2)}{\Gamma(3-d/2)\Gamma(d/2-1)^2}\frac{d^dl}{i\pi^{d/2}}\,,
\end{align}
so that $M_{3,1}(4-2\epsilon)=\epsilon^{-4}$. This choice corresponds to the normalization of Refs.~\cite{Baikov:2010hf,Lee:2011jf}.

Below we list our results ordered by the complexity level (c.l.) \cite{Lee:2010cga}.

\textbf{Integrals with c.l.=1}
\begin{align}
&M_{2,1}+\frac{(3-7 \epsilon ) (3-5 \epsilon ) (4-5 \epsilon ) (3-4 \epsilon )}{(1-3 \epsilon ) (1-2 \epsilon )^3}M_{0,1}
-\frac{(3-5 \epsilon ) (2-3 \epsilon ) (1-\epsilon )}{2 (1-4 \epsilon ) (1-2 \epsilon )^2}M_{1,3}
\nonumber\\
&
=\frac{1}{1-4 \epsilon }
\Biggl\{
   -\frac{5}{48 \epsilon ^3}-\frac{19 \zeta _3}{12}-\frac{19 \zeta _4 \epsilon }{8}-\frac{341 \zeta _5 \epsilon ^2}{4}-\biggl(-\frac{493 \zeta _3^2}{6}+\frac{1255 \zeta _6}{6}\biggr) \epsilon ^3
   \nonumber\\
   &-\biggl(-\frac{493}{2} \zeta _3 \zeta _4+\frac{16619 \zeta _7}{8}\biggr) \epsilon ^4
   - \biggl(-2402 \zeta _3 \zeta _5+\frac{61523 \zeta _8}{16}+\frac{1215 \zeta _{2,6}}{2}\biggr)\epsilon ^5
   \nonumber\\
   &-\biggl(\frac{13166 \zeta _3^3}{9}-7248 \zeta _4 \zeta _5-\frac{25895 \zeta _3 \zeta _6}{3}+\frac{3258785 \zeta _9}{72}\biggr) \epsilon ^6
   -\biggl(6583 \zeta _3^2 \zeta _4
   \nonumber\\
   &-\frac{4293369 \zeta _5^2}{112}-\frac{7884283 \zeta _3 \zeta _7}{112}+\frac{28842885 \zeta _{10}}{224}-\frac{242325 \zeta _{3,7}}{112}\biggr)\epsilon ^7
   - \biggl(49534 \zeta _3^2 \zeta _5
   \nonumber\\
   &-\frac{453725 \zeta _5 \zeta _6}{2}-\frac{1981671 \zeta _4 \zeta _7}{16}-182183 \zeta _3 \zeta _8+173160 \zeta _2 \zeta _9+\frac{90139333 \zeta _{11}}{128}
   \nonumber\\
   &
   -19080 \zeta _3 \zeta _{2,6}+9360 \zeta _{2,1,8}\biggr)\epsilon ^8-\biggl(-\frac{290335 \zeta _3^4}{18}+\frac{536919}{2} \zeta _3 \zeta _4 \zeta _5
   \nonumber\\
   &+\frac{2074005}{14} \zeta _2 \zeta _5^2+\frac{2107405}{12} \zeta _3^2 \zeta _6+\frac{2108145}{7} \zeta _2 \zeta _3 \zeta _7-\frac{4328161 \zeta _5 \zeta _7}{2}
   \nonumber\\
   &-\frac{146898725 \zeta _3 \zeta _9}{72}+\frac{1792356175927 \zeta _{12}}{530688}-\frac{59715}{2} \zeta _4 \zeta _{2,6}-\frac{76815}{7} \zeta _2 \zeta _{3,7}
   \nonumber\\
   &+\frac{270635 \zeta _{3,9}}{24}+17070 \zeta _{2,1,1,8}\biggr)\epsilon ^9+O\left(\epsilon ^{10}\right)
\Biggr\}\,,
\end{align}
\begin{align}
&M_{2,2}-\frac{4 (3-5 \epsilon ) (4-5 \epsilon ) (3-4 \epsilon )}{(1-3 \epsilon ) (1-2 \epsilon )^2}M_{0,1}
+\frac{(2-3 \epsilon )^2}{(2-5 \epsilon ) (1-2 \epsilon )}M_{1,2}
\nonumber\\
&
=
\frac{2 (1-2 \epsilon )}{(1-5 \epsilon ) (2-5 \epsilon )}
\Biggl\{
   -\frac{1}{4 \epsilon ^3}+10 \zeta _3+15 \zeta _4 \epsilon +185 \zeta _5 \epsilon ^2+\biggl(-204 \zeta _3^2+\frac{875 \zeta _6}{2}\biggr) \epsilon ^3
   \nonumber\\
   &+\biggl(-612 \zeta _3 \zeta _4+\frac{13157 \zeta _7}{4}\biggr) \epsilon ^4
   + \biggl(-6830 \zeta _3 \zeta _5+\frac{67767 \zeta _8}{8}+243 \zeta _{2,6}\biggr)\epsilon ^5
   \nonumber\\
   &+\biggl(\frac{8644 \zeta _3^3}{3}-11703 \zeta _4 \zeta _5-17270 \zeta _3 \zeta _6+\frac{735227 \zeta _9}{12}\biggr) \epsilon ^6
   -\biggl(-12966 \zeta _3^2 \zeta _4
   \nonumber\\
   &+\frac{3770679 \zeta _5^2}{56}+\frac{7170073 \zeta _3 \zeta _7}{56}-\frac{20316339 \zeta _{10}}{112}+\frac{40851 \zeta _{3,7}}{56}\biggr)\epsilon ^7
   \nonumber\\
   &
   + \biggl(133222 \zeta _3^2 \zeta _5-\frac{659003 \zeta _5 \zeta _6}{2}-\frac{1644819 \zeta _4 \zeta _7}{8}-\frac{1341627 \zeta _3 \zeta _8}{4}+16317 \zeta _2 \zeta _9
   \nonumber\\
   &+\frac{75373337 \zeta _{11}}{64}-9144 \zeta _3 \zeta _{2,6}+882 \zeta _{2,1,8}\biggr)\epsilon ^8
   - \biggl(\frac{94907 \zeta _3^4}{3}-459003 \zeta _3 \zeta _4 \zeta _5
   \nonumber\\
   &-\frac{143613}{7} \zeta _2 \zeta _5^2-\frac{707359}{2} \zeta _3^2 \zeta _6-\frac{291954}{7} \zeta _2 \zeta _3 \zeta _7+\frac{4983229 \zeta _5 \zeta _7}{2}+\frac{9870127 \zeta _3 \zeta _9}{4}
   \nonumber\\
   &-\frac{976579279975 \zeta _{12}}{265344}+13266 \zeta _4 \zeta _{2,6}+\frac{10638}{7} \zeta _2 \zeta _{3,7}+\frac{19445 \zeta _{3,9}}{12}
   \nonumber\\
   &-2364 \zeta _{2,1,1,8}\biggr)\epsilon ^9+O\left(\epsilon ^{10}\right)
\Biggr\}\,,
\end{align}
\begin{align}
&M_{2,6}+\frac{4 (3-5 \epsilon ) (4-5 \epsilon ) (3-4 \epsilon )}{(1-3 \epsilon )^2 (1-2 \epsilon )}M_{0,1}
=\frac{2 (1-2 \epsilon )^2}{(1-5 \epsilon ) (2-5 \epsilon ) (1-3 \epsilon )}
\Biggl\{
   -\frac{1}{8 \epsilon ^3}
   \nonumber\\
   &+2 \zeta _3+3 \zeta _4 \epsilon +24 \zeta _5 \epsilon ^2+\biggl(\frac{49 \zeta _3^2}{2}+55 \zeta _6\biggr) \epsilon ^3+\biggl(\frac{147 \zeta _3 \zeta _4}{2}+\frac{2475 \zeta _7}{4}\biggr) \epsilon ^4
   \nonumber\\
   &
   + \biggl(507 \zeta _3 \zeta _5+\frac{48531 \zeta _8}{16}-243 \zeta _{2,6}\biggr)\epsilon ^5+\biggl(-\frac{3875 \zeta _3^3}{3}+\frac{4437 \zeta _4 \zeta _5}{2}+2360 \zeta _3 \zeta _6
   \nonumber\\
   &
   +\frac{49652 \zeta _9}{3}\biggr) \epsilon ^6+\biggl(-\frac{11625}{2} \zeta _3^2 \zeta _4+\frac{327123 \zeta _5^2}{28}+\frac{239661 \zeta _3 \zeta _7}{28}+\frac{4104945 \zeta _{10}}{56}
   \nonumber\\
   &
   +\frac{17253 \zeta _{3,7}}{28}\biggr)\epsilon ^7+ \biggl(-45204 \zeta _3^2 \zeta _5+\frac{269691 \zeta _5 \zeta _6}{4}+\frac{78381 \zeta _4 \zeta _7}{4}-\frac{142251 \zeta _3 \zeta _8}{8}
   \nonumber\\
   &-\frac{80919 \zeta _2 \zeta _9}{2}+\frac{7592883 \zeta _{11}}{16}+10449 \zeta _3 \zeta _{2,6}-2187 \zeta _{2,1,8}\biggr)\epsilon ^8
   +\biggl(\frac{143995 \zeta _3^4}{6}
   \nonumber\\
   &-188829 \zeta _3 \zeta _4 \zeta _5-\frac{118098}{7} \zeta _2 \zeta _5^2-149047 \zeta _3^2 \zeta _6-\frac{240084}{7} \zeta _2 \zeta _3 \zeta _7+\frac{727515 \zeta _5 \zeta _7}{2}
   \nonumber\\
   &+\frac{161044 \zeta _3 \zeta _9}{3}+\frac{79499888185 \zeta _{12}}{44224}+17496 \zeta _4 \zeta _{2,6}+\frac{8748}{7} \zeta _2 \zeta _{3,7}+675 \zeta _{3,9}
   \nonumber\\
   &-1944 \zeta _{2,1,1,8}\biggr)\epsilon ^9+O\left(\epsilon ^{10}\right)
\Biggr\}\,,
\end{align}
\begin{align}
&M_{2,7}-\frac{(3-5 \epsilon ) (4-5 \epsilon ) (3-4 \epsilon ) \left(7-41 \epsilon +58 \epsilon ^2\right)}{(1-4 \epsilon ) (1-3 \epsilon
   ) (2-3 \epsilon ) (1-2 \epsilon )^2}M_{0,1}
=
\frac{2 (1-2 \epsilon )}{(1-4 \epsilon ) (2-3 \epsilon )}
   \nonumber\\
   &\times
\Biggl\{
   \frac{1}{48 \epsilon ^3}
   +\frac{7 \zeta _3}{6}+\frac{7 \zeta _4 \epsilon }{4}+\frac{221 \zeta _5 \epsilon ^2}{4}+\biggl(-\frac{145 \zeta _3^2}{3}+\frac{3245 \zeta _6}{24}\biggr) \epsilon ^3+\biggl(\frac{11289 \zeta _7}{8}
   \nonumber\\
   &
   -145 \zeta _3 \zeta _4\biggr) \epsilon ^4+\biggl(-875 \zeta _3 \zeta _5+\frac{7441 \zeta _8}{4}+\frac{1215 \zeta _{2,6}}{2}\biggr)\epsilon ^5 +\biggl(\frac{8404 \zeta _3^3}{9}-\frac{9915 \zeta _4 \zeta _5}{2}
   \nonumber\\
   &-\frac{14950 \zeta _3 \zeta _6}{3}+\frac{286295 \zeta _9}{9}\biggr) \epsilon ^6
   +\biggl(4202 \zeta _3^2 \zeta _4-\frac{600375 \zeta _5^2}{28}-\frac{1124475 \zeta _3 \zeta _7}{28}
    \nonumber\\
   &
   +\frac{4628241 \zeta _{10}}{56}-\frac{71685 \zeta _{3,7}}{28}\biggr)\epsilon ^7+\biggl(17701 \zeta _3^2 \zeta _5-\frac{618655 \zeta _5 \zeta _6}{4}-82305 \zeta _4 \zeta _7
   \nonumber\\
   &-96139 \zeta _3 \zeta _8+\frac{404595 \zeta _2 \zeta _9}{2}+\frac{3052633 \zeta _{11}}{8}-20655 \zeta _3 \zeta _{2,6}+10935 \zeta _{2,1,8}\biggr)\epsilon ^8
   \nonumber\\
   &
   +\biggl(-\frac{105418 \zeta _3^4}{9}+129648 \zeta _3 \zeta _4 \zeta _5+\frac{590490}{7} \zeta _2 \zeta _5^2+\frac{323765}{3} \zeta _3^2 \zeta _6
   \nonumber\\
   &+\frac{1200420}{7} \zeta _2 \zeta _3 \zeta _7-1237050 \zeta _5 \zeta _7-\frac{10667105 \zeta _3 \zeta _9}{9}+\frac{36344694233 \zeta _{12}}{16584}
   \nonumber\\
   &-40095 \zeta _4 \zeta _{2,6}-\frac{43740}{7} \zeta _2 \zeta _{3,7}-11475 \zeta _{3,9}+9720 \zeta _{2,1,1,8}\biggr)\epsilon ^9+O\left(\epsilon ^{10}\right)
\Biggr\}\,,
\end{align}
\begin{align}
&M_{3,2}+\frac{4 (3-4 \epsilon ) (2-3 \epsilon )}{(1-3 \epsilon ) (1-2 \epsilon )}M_{1,1}
=
\frac{1-2 \epsilon }{1-3 \epsilon }
\Biggl\{
   \frac{1}{3 \epsilon ^4}+\frac{14 \zeta _3}{3 \epsilon }+7 \zeta _4+126 \zeta _5 \epsilon
   \nonumber\\
   &+\biggl(-\frac{226 \zeta _3^2}{3}+\frac{910 \zeta _6}{3}\biggr) \epsilon ^2+\biggl(-226 \zeta _3 \zeta _4+1960 \zeta _7\biggr) \epsilon ^3
   +\epsilon ^4 \biggl(-612 \zeta _3 \zeta _5
   \nonumber\\
   &+\frac{11851 \zeta _8}{4}+648 \zeta _{2,6}\biggr)+\biggl(\frac{5260 \zeta _3^3}{9}-4806 \zeta _4 \zeta _5-\frac{13180 \zeta _3 \zeta _6}{3}+\frac{247094 \zeta _9}{9}\biggr) \epsilon ^5
   \nonumber\\
   &
   + \biggl(2630 \zeta _3^2 \zeta _4-6912 \zeta _5^2-11926 \zeta _3 \zeta _7+\frac{843327 \zeta _{10}}{20}+5751 \zeta _{2,8}\biggr)\epsilon ^6
   + \biggl(8244 \zeta _3^2 \zeta _5
   \nonumber\\
   &-93294 \zeta _5 \zeta _6-50694 \zeta _4 \zeta _7-\frac{115061 \zeta _3 \zeta _8}{2}+107892 \zeta _2 \zeta _9+\frac{417459 \zeta _{11}}{2}
   \nonumber\\
   &-8424 \zeta _3 \zeta _{2,6}+5832 \zeta _{2,1,8}\biggr)\epsilon ^7
   + \biggl(-\frac{25330 \zeta _3^4}{9}+50004 \zeta _3 \zeta _4 \zeta _5+58320 \zeta _2 \zeta _5^2
   \nonumber\\
   &+\frac{122876}{3} \zeta _3^2 \zeta _6+111456 \zeta _2 \zeta _3 \zeta _7-524340 \zeta _5 \zeta _7-\frac{4054892 \zeta _3 \zeta _9}{9}-17496 \zeta _4 \zeta _{2,6}
   \nonumber\\
   &+\frac{30212886835 \zeta _{12}}{33168}+11664 \zeta _2 \zeta _{2,8}+8100 \zeta _{2,10}+5184 \zeta _{2,1,1,8}\biggr)\epsilon ^8+O\left(\epsilon ^9\right)
\Biggr\}\,,
\end{align}
\begin{align}
&M_{3,3}+\frac{4 (2-5 \epsilon ) (3-5 \epsilon )}{(1-4 \epsilon ) (1-3 \epsilon )}M_{1,4}
=
\frac{1-2 \epsilon }{1-4 \epsilon }
\Biggl\{
   \frac{1}{6 \epsilon ^4}+\frac{31 \zeta _3}{3 \epsilon }+\frac{31 \zeta _4}{2}+449 \zeta _5 \epsilon
   \nonumber\\
   &+\biggl(\frac{3290 \zeta _6}{3}-\frac{983 \zeta _3^2}{3}\biggr) \epsilon ^2+\biggl(11338 \zeta _7-983 \zeta _3 \zeta _4\biggr) \epsilon ^3
   +\biggl(4860 \zeta _{2,6}-3914 \zeta _3 \zeta _5
   \nonumber\\
   &+\frac{121261 \zeta _8}{8}\biggr)\epsilon ^4 +\biggl(\frac{47918 \zeta _3^3}{9}-35031 \zeta _4 \zeta _5-\frac{97340 \zeta _3 \zeta _6}{3}+\frac{2293555 \zeta _9}{9}\biggr) \epsilon ^5
   \nonumber\\&
   + \biggl(23959 \zeta _3^2 \zeta _4-\frac{1069773 \zeta _5^2}{7}-\frac{1753546 \zeta _3 \zeta _7}{7}+\frac{4720473 \zeta _{10}}{7}-\frac{143370 \zeta _{3,7}}{7}\biggr)\epsilon ^6
   \nonumber\\
   &
   + \biggl(93802 \zeta _3^2 \zeta _5-1151470 \zeta _5 \zeta _6-552282 \zeta _4 \zeta _7-\frac{2732573 \zeta _3 \zeta _8}{4}+1618380 \zeta _2 \zeta _9
   \nonumber\\
   &+3055286 \zeta _{11}-136080 \zeta _3 \zeta _{2,6}+87480 \zeta _{2,1,8}\biggr)\epsilon ^7
   + \biggl(718806 \zeta _3 \zeta _4 \zeta _5-\frac{499487 \zeta _3^4}{9}
   \nonumber\\
   &+\frac{4723920}{7} \zeta _2 \zeta _5^2+\frac{1842460}{3} \zeta _3^2 \zeta _6+\frac{9603360}{7} \zeta _2 \zeta _3 \zeta _7-9290256 \zeta _5 \zeta _7
   \nonumber\\
   &-\frac{71396590 \zeta _3 \zeta _9}{9}+\frac{1180748875591 \zeta _{12}}{66336}-277020 \zeta _4 \zeta _{2,6}-\frac{349920}{7} \zeta _2 \zeta _{3,7}
   \nonumber\\
   &-91800 \zeta _{3,9}+77760 \zeta _{2,1,1,8}\biggr)\epsilon ^8+O\left(\epsilon ^9\right)
\Biggr\}\,,
\end{align}
\begin{align}
&M_{3,4}+\frac{2 (2-5 \epsilon ) (3-5 \epsilon ) (1-2 \epsilon )}{(1-4 \epsilon ) (1-3 \epsilon )^2}M_{1,4}
=
\frac{(1-2 \epsilon )^2}{(1-4 \epsilon ) (1-3 \epsilon )}
\Biggl\{
   \frac{1}{12 \epsilon ^4}+\frac{25 \zeta _3}{6 \epsilon }
   \nonumber\\
   &+\frac{25 \zeta _4}{4}+\frac{465 \zeta _5 \epsilon }{2}+\biggl(-\frac{1247 \zeta _3^2}{6}+\frac{3425 \zeta _6}{6}\biggr) \epsilon ^2+\biggl(-\frac{1247}{2} \zeta _3 \zeta _4+\frac{12503 \zeta _7}{2}\biggr) \epsilon ^3
   \nonumber\\
   &+\biggl(1944 \zeta _{2,6}-5895 \zeta _3 \zeta _5+\frac{182497 \zeta _8}{16}\biggr)\epsilon ^4 +\biggl(\frac{37081 \zeta _3^3}{9}-\frac{41013 \zeta _4 \zeta _5}{2}
   \nonumber\\
   &-\frac{70255 \zeta _3 \zeta _6}{3}+\frac{2619709 \zeta _9}{18}\biggr) \epsilon ^5
   + \biggl(\frac{37081}{2} \zeta _3^2 \zeta _4-\frac{1520763 \zeta _5^2}{14}-\frac{1391503 \zeta _3 \zeta _7}{7}
   \nonumber\\
   &+\frac{11712423 \zeta _{10}}{28}-\frac{55656 \zeta _{3,7}}{7}\biggr)\epsilon ^6
   + \biggl(144561 \zeta _3^2 \zeta _5-680047 \zeta _5 \zeta _6-\frac{741363 \zeta _4 \zeta _7}{2}
   \nonumber\\
   &-\frac{4252007 \zeta _3 \zeta _8}{8}+605616 \zeta _2 \zeta _9+\frac{4622343 \zeta _{11}}{2}-59952 \zeta _3 \zeta _{2,6}+32736 \zeta _{2,1,8}\biggr)\epsilon ^7
   \nonumber\\
   &
   +\biggl(-\frac{870911 \zeta _3^4}{18}+783699 \zeta _3 \zeta _4 \zeta _5+\frac{3297024}{7} \zeta _2 \zeta _5^2+\frac{1557073}{3} \zeta _3^2 \zeta _6
   \nonumber\\
   &+\frac{6702592}{7} \zeta _2 \zeta _3 \zeta _7-\frac{19417073 \zeta _5 \zeta _7}{3}-\frac{55088251 \zeta _3 \zeta _9}{9}+\frac{4433462543495 \zeta _{12}}{398016}
   \nonumber\\
   &-98328 \zeta _4 \zeta _{2,6}-\frac{244224}{7} \zeta _2 \zeta _{3,7}+\frac{94000 \zeta _{3,9}}{9}+54272 \zeta _{2,1,1,8}\biggr)\epsilon ^8 +O\left(\epsilon ^9\right)
\Biggr\}\,,
\end{align}
\begin{align}
&M_{3,6}=
\frac{(1-2 \epsilon )^3}{1-5 \epsilon }
\Biggl\{
   \frac{5 \zeta _5}{\epsilon }+\biggl(\frac{25 \zeta _6}{2}-7 \zeta _3^2\biggr)+\biggl(\frac{127 \zeta _7-21 \zeta _3 \zeta _4}{2}\biggr) \epsilon
   - \biggl(-994 \zeta _3 \zeta _5
   \nonumber\\
   &+\frac{14595 \zeta _8}{8}-486 \zeta _{2,6}\biggr)\epsilon ^2-\biggl(-\frac{1742 \zeta _3^3}{3}+1425 \zeta _4 \zeta _5-90 \zeta _3 \zeta _6-\frac{346 \zeta _9}{3}\biggr) \epsilon ^3
   \nonumber\\
   &
   -\biggl(-2613 \zeta _3^2 \zeta _4+\frac{32958 \zeta _5^2}{7}-\frac{73231 \zeta _3 \zeta _7}{7}+\frac{124308 \zeta _{10}}{7}+\frac{11799 \zeta _{3,7}}{7}\biggr)\epsilon ^4
   \nonumber\\
   &
   - \biggl(37843 \zeta _5 \zeta _6-334 \zeta _3^2 \zeta _5+39900 \zeta _4 \zeta _7-\frac{542969 \zeta _{11}}{4}+100566 \zeta _2 \zeta _9+5436 \zeta _{2,1,8}
   \nonumber\\
   &-\frac{383805 \zeta _3 \zeta _8}{4}+8172 \zeta _3 \zeta _{2,6}\biggr)\epsilon ^5
   - \biggl(13891 \zeta _3^4-135750 \zeta _3 \zeta _4 \zeta _5-\frac{425736}{7} \zeta _2 \zeta _5^2
   \nonumber\\
   &-47134 \zeta _3^2 \zeta _6-\frac{865488}{7} \zeta _2 \zeta _3 \zeta _7+410096 \zeta _5 \zeta _7-\frac{271438 \zeta _3 \zeta _9}{3}+\frac{23122877963 \zeta _{12}}{66336}
   \nonumber\\
   &-1152 \zeta _4 \zeta _{2,6}+\frac{31536}{7} \zeta _2 \zeta _{3,7}+\frac{14840 \zeta _{3,9}}{3}-7008 \zeta _{2,1,1,8}\biggr)\epsilon ^6+O\left(\epsilon ^7\right)
\Biggr\}\,,
\end{align}

\textbf{Integrals with c.l.=2}
\begin{align}
&M_{3,5}=
\frac{(1-2 \epsilon )^3}{(1-5 \epsilon ) (1-4 \epsilon )}
\Biggl\{
   \frac{\zeta _3}{2 \epsilon ^2}+\frac{3 \zeta _4}{4 \epsilon }-\frac{23 \zeta _5}{2}-\biggl(-\frac{29 \zeta _3^2}{2}+30 \zeta _6\biggr) \epsilon
   \nonumber\\
   &-\biggl(-\frac{87}{2} \zeta _3 \zeta _4+\frac{1105 \zeta _7}{4}\biggr) \epsilon ^2
   + \biggl(1639 \zeta _3 \zeta _5-\frac{29043 \zeta _8}{16}+243 \zeta _{2,6}\biggr)\epsilon ^3-\biggl(967 \zeta _3^3
   \nonumber\\
   &-\frac{2001 \zeta _4 \zeta _5}{2}-2810 \zeta _3 \zeta _6+5144 \zeta _9\biggr) \epsilon ^4
   + \biggl(-\frac{8703}{2} \zeta _3^2 \zeta _4+\frac{35235 \zeta _5^2}{2}+\frac{43361 \zeta _3 \zeta _7}{2}
   \nonumber\\
   &-\frac{105813 \zeta _{10}}{4}-1296 \zeta _{3,7}\biggr)\epsilon ^5
   - \biggl(86492 \zeta _3^2 \zeta _5-\frac{288269 \zeta _5 \zeta _6}{4}+\frac{14163 \zeta _4 \zeta _7}{2}
   \nonumber\\
   &-\frac{917679 \zeta _3 \zeta _8}{8}+\frac{118881 \zeta _2 \zeta _9}{2}-\frac{12109 \zeta _{11}}{32}+10881 \zeta _3 \zeta _{2,6}+3213 \zeta _{2,1,8}\biggr)\epsilon ^6
   \nonumber\\
   &
   + \biggl(\frac{120277 \zeta _3^4}{6}-256497 \zeta _3 \zeta _4 \zeta _5-\frac{1055592}{7} \zeta _2 \zeta _5^2-137128 \zeta _3^2 \zeta _6-\frac{2145936}{7} \zeta _2 \zeta _3 \zeta _7
   \nonumber\\
   &+\frac{2500975 \zeta _5 \zeta _7}{2}+\frac{1911321 \zeta _3 \zeta _9}{2}-\frac{55562910103 \zeta _{12}}{66336}-24534 \zeta _4 \zeta _{2,6}
   \nonumber\\
   &+\frac{78192}{7} \zeta _2 \zeta _{3,7}-\frac{291485 \zeta _{3,9}}{6}-17376 \zeta _{2,1,1,8}\biggr)\epsilon ^7+O\left(\epsilon ^8\right)
\Biggr\}\,,
\end{align}
\begin{align}
&M_{4,1}=
(1-2 \epsilon )^2
\Biggl\{
   \frac{20 \zeta _5}{\epsilon }+\biggl(-22 \zeta _3^2+50 \zeta _6\biggr)+\biggl(-66 \zeta _3 \zeta _4+\frac{4685 \zeta _7}{8}\biggr) \epsilon
   \nonumber\\
   &
   -\biggl(-4835 \zeta _3 \zeta _5+\frac{119987 \zeta _8}{16}-\frac{4563 \zeta _{2,6}}{2}\biggr)\epsilon ^2 +\biggl(860 \zeta _3^3-\frac{12873 \zeta _4 \zeta _5}{2}+790 \zeta _3 \zeta _6
   \nonumber\\
   &+\frac{102457 \zeta _9}{8}\biggr) \epsilon ^3- \biggl(-3870 \zeta _3^2 \zeta _4-\frac{622205 \zeta _5^2}{112}-\frac{3987915 \zeta _3 \zeta _7}{112}+\frac{13703985 \zeta _{10}}{224}
   \nonumber\\
   &
   +\frac{1210959 \zeta _{3,7}}{112}\biggr)\epsilon ^4+ \biggl(12996 \zeta _{2,1,8}-122760 \zeta _3^2 \zeta _5-\frac{288509 \zeta _5 \zeta _6}{2}-\frac{2362023 \zeta _4 \zeta _7}{16}
   \nonumber\\
   &+\frac{2536537 \zeta _3 \zeta _8}{8}+240426 \zeta _2 \zeta _9-\frac{14723695 \zeta _{11}}{128}-71613 \zeta _3 \zeta _{2,6}\biggr)\epsilon ^5
   - \biggl(\frac{121663 \zeta _3^4}{6}
   \nonumber\\
   &+\frac{415317}{2} \zeta _3 \zeta _4 \zeta _5+\frac{3519855}{14} \zeta _2 \zeta _5^2-\frac{335005}{12} \zeta _3^2 \zeta _6+\frac{3577795}{7} \zeta _2 \zeta _3 \zeta _7
   \nonumber\\
   &-\frac{4191530 \zeta _5 \zeta _7}{3}-\frac{140309257 \zeta _3 \zeta _9}{72}+\frac{3406738440035 \zeta _{12}}{1592064}+148641 \zeta _4 \zeta _{2,6}
   \nonumber\\
   &-\frac{130365}{7} \zeta _2 \zeta _{3,7}+\frac{14194343 \zeta _{3,9}}{72}+28970 \zeta _{2,1,1,8}\biggr)\epsilon ^6+O\left(\epsilon ^7\right)
\Biggr\}\,,
\end{align}
\begin{align}
&M_{4,2}=
(1-2 \epsilon )^2
\Biggl\{
   \frac{20 \zeta _5}{\epsilon }+\biggl(8 \zeta _3^2+50 \zeta _6\biggr)+\biggl(24 \zeta _3 \zeta _4+520 \zeta _7\biggr) \epsilon
   + \biggl(-4120 \zeta _3 \zeta _5
   \nonumber\\
   &+6958 \zeta _8-1296 \zeta _{2,6}\biggr)\epsilon ^2+\biggl(-320 \zeta _3^3+1596 \zeta _4 \zeta _5-3860 \zeta _3 \zeta _6+\frac{23753 \zeta _9}{2}\biggr) \epsilon ^3
   \nonumber\\
   &
   + \biggl(-1440 \zeta _3^2 \zeta _4-\frac{691465 \zeta _5^2}{28}-\frac{1315395 \zeta _3 \zeta _7}{28}+\frac{5181705 \zeta _{10}}{56}+\frac{151659 \zeta _{3,7}}{28}\biggr)\epsilon ^4
   \nonumber\\
   &
   + \biggl(85950 \zeta _3^2 \zeta _5-\frac{63389 \zeta _5 \zeta _6}{2}+\frac{97263 \zeta _4 \zeta _7}{4}-234601 \zeta _3 \zeta _8-154179 \zeta _2 \zeta _9
   \nonumber\\
   &+\frac{16056145 \zeta _{11}}{32}+34902 \zeta _3 \zeta _{2,6}-8334 \zeta _{2,1,8}\biggr)\epsilon ^5
   + \biggl(3682 \zeta _3^4-35946 \zeta _3 \zeta _4 \zeta _5
   \nonumber\\
   &-\frac{1844370}{7} \zeta _2 \zeta _5^2+80045 \zeta _3^2 \zeta _6-\frac{3749460}{7} \zeta _2 \zeta _3 \zeta _7+\frac{387745 \zeta _5 \zeta _7}{2}+\frac{399371 \zeta _3 \zeta _9}{6}
   \nonumber\\
   &+\frac{153248180675 \zeta _{12}}{132672}+42084 \zeta _4 \zeta _{2,6}+\frac{136620}{7} \zeta _2 \zeta _{3,7}+\frac{13691 \zeta _{3,9}}{6}
   \nonumber\\
   &-30360 \zeta _{2,1,1,8}\biggr)\epsilon ^6+O\left(\epsilon ^7\right)
\Biggr\}\,,
\end{align}
\begin{align}
&M_{4,3}=
\frac{(1-2 \epsilon )^3}{1+3 \epsilon }
\Biggl\{
   -\frac{5 \zeta _5}{\epsilon }-\biggl(17 \zeta _3^2+\frac{25 \zeta _6}{2}\biggr)-\biggl(51 \zeta _3 \zeta _4+\frac{225 \zeta _7}{2}\biggr) \epsilon
   \nonumber\\
   &
   -\biggl(\frac{40201 \zeta _8}{8}-3030 \zeta _3 \zeta _5-1134 \zeta _{2,6}\biggr)\epsilon ^2 -\biggl(-\frac{2690 \zeta _3^3}{3}+2259 \zeta _4 \zeta _5-1990 \zeta _3 \zeta _6
   \nonumber\\&
   +\frac{22009 \zeta _9}{9}\biggr) \epsilon ^3+\biggl(4035 \zeta _3^2 \zeta _4+\frac{125425 \zeta _5^2}{14}+\frac{289795 \zeta _3 \zeta _7}{14}-\frac{1059305 \zeta _{10}}{28}
   \nonumber\\
   &
   -\frac{42513 \zeta _{3,7}}{14}\biggr)\epsilon ^4+ \biggl(-99740 \zeta _3^2 \zeta _5-\frac{30257 \zeta _5 \zeta _6}{6}-\frac{41491 \zeta _4 \zeta _7}{2}+\frac{930921 \zeta _3 \zeta _8}{4}
   \nonumber\\
   &+94091 \zeta _2 \zeta _9-\frac{1536475 \zeta _{11}}{8}-48178 \zeta _3 \zeta _{2,6}+5086 \zeta _{2,1,8}\biggr)\epsilon ^5
   + \biggl(-\frac{204427 \zeta _3^4}{9}
   \nonumber\\
   &-35626 \zeta _3 \zeta _4 \zeta _5+\frac{234900}{7} \zeta _2 \zeta _5^2-\frac{269950}{9} \zeta _3^2 \zeta _6+\frac{1432600}{21} \zeta _2 \zeta _3 \zeta _7+\frac{1200535 \zeta _5 \zeta _7}{9}
   \nonumber\\
   &+\frac{5297692 \zeta _3 \zeta _9}{27}-\frac{94566489365 \zeta _{12}}{597024}-76996 \zeta _4 \zeta _{2,6}-\frac{17400}{7} \zeta _2 \zeta _{3,7}-\frac{404498 \zeta _{3,9}}{27}
   \nonumber\\
   &+\frac{11600}{3} \zeta _{2,1,1,8}\biggr)\epsilon ^6+O\left(\epsilon ^7\right)
\Biggr\}\,,
\end{align}
\begin{align}
&M_{4,4}=
(1-2 \epsilon )^3
\Biggl\{
   \frac{441 \zeta _7}{8}
   +\epsilon  \biggl(-216 \zeta _3 \zeta _5+\frac{5733 \zeta _8}{16}-\frac{81 \zeta _{2,6}}{2}\biggr)+\biggl(-267 \zeta _3^3
   \nonumber\\
   &-81 \zeta _4 \zeta _5-\frac{675 \zeta _3 \zeta _6}{2}+\frac{4583 \zeta _9}{2}\biggr) \epsilon ^2
   +\biggl(-\frac{2403}{2} \zeta _3^2 \zeta _4-\frac{502287 \zeta _5^2}{56}-\frac{7731 \zeta _3 \zeta _7}{56}
   \nonumber\\
   &+\frac{1324935 \zeta _{10}}{112}+\frac{18441 \zeta _{3,7}}{56}\biggr)\epsilon ^3
   + \biggl(-\frac{24315}{2} \zeta _3^2 \zeta _5-\frac{358023 \zeta _5 \zeta _6}{8}+\frac{139401 \zeta _4 \zeta _7}{8}\nonumber\\
   &-\frac{59895 \zeta _3 \zeta _8}{4}+\frac{232767 \zeta _2 \zeta _9}{4}-\frac{402081 \zeta _{11}}{32}-\frac{621}{2} \zeta _3 \zeta _{2,6}+\frac{6291}{2} \zeta _{2,1,8}\biggr)\epsilon ^4
   \nonumber\\
   &- \biggl(-6023 \zeta _3^4+6660 \zeta _3 \zeta _4 \zeta _5-\frac{650997}{7} \zeta _2 \zeta _5^2+40507 \zeta _3^2 \zeta _6-\frac{1323426}{7} \zeta _2 \zeta _3 \zeta _7
   \nonumber\\
   &+\frac{1750957 \zeta _5 \zeta _7}{2}+\frac{964778 \zeta _3 \zeta _9}{3}-\frac{104287641323 \zeta _{12}}{132672}-2853 \zeta _4 \zeta _{2,6}+\frac{48222}{7} \zeta _2 \zeta _{3,7}
   \nonumber\\
   &-\frac{190175 \zeta _{3,9}}{6}-10716 \zeta _{2,1,1,8}\biggr)\epsilon ^5+O\left(\epsilon ^6\right)
\Biggr\}\,,
\end{align}
\begin{align}
&M_{5,2}(4-2\epsilon)
=
(1-2 \epsilon )^2
\Biggl\{
   \frac{20 \zeta _5}{\epsilon }+\biggl(68 \zeta _3^2+50 \zeta _6\biggr)+\biggl(204 \zeta _3 \zeta _4+450 \zeta _7\biggr) \epsilon
   \nonumber\\
   &
   +\biggl(-11520 \zeta _3 \zeta _5+\frac{40201 \zeta _8}{2}-4536 \zeta _{2,6}\biggr)\epsilon ^2 +\biggl(9936 \zeta _4 \zeta _5-\frac{4640 \zeta _3^3}{3}-6460 \zeta _3 \zeta _6
   \nonumber\\
   &+\frac{88036 \zeta _9}{9}\biggr) \epsilon ^3
   +\biggl(-6960 \zeta _3^2 \zeta _4-73022 \zeta _5^2-142178 \zeta _3 \zeta _7+\frac{6483199 \zeta _{10}}{20}
   \nonumber\\
   &-42513 \zeta _{2,8}\biggr)\epsilon ^4
   + \biggl(83120 \zeta _3^2 \zeta _5+\frac{227014 \zeta _5 \zeta _6}{3}+103232 \zeta _4 \zeta _7-317196 \zeta _3 \zeta _8
   \nonumber\\
   &-376364 \zeta _2 \zeta _9+\frac{1536475 \zeta _{11}}{2}+56632 \zeta _3 \zeta _{2,6}-20344 \zeta _{2,1,8}\biggr)\epsilon ^5
   + \biggl(\frac{124708 \zeta _3^4}{9}
   \nonumber\\
   &+11464 \zeta _3 \zeta _4 \zeta _5-174000 \zeta _2 \zeta _5^2-\frac{40700}{9} \zeta _3^2 \zeta _6-\frac{997600}{3} \zeta _2 \zeta _3 \zeta _7-\frac{2338736 \zeta _5 \zeta _7}{3}
   \nonumber\\
   &-\frac{26211464 \zeta _3 \zeta _9}{27}+\frac{9482746239 \zeta _{12}}{2764}+103864 \zeta _4 \zeta _{2,6}-34800 \zeta _2 \zeta _{2,8}-\frac{808996 \zeta _{2,10}}{3}
   \nonumber\\
   &-\frac{46400}{3} \zeta _{2,1,1,8}\biggr)\epsilon ^6+O\left(\epsilon ^7\right)
\Biggr\}\,,
\end{align}

\textbf{Integrals with c.l.=3}
\begin{align}
&M_{4,5}(4-2\epsilon)
=
\frac{(1-2 \epsilon )^3}{1-6 \epsilon }
\Biggl\{
   36 \zeta _3^2-\biggl(-108 \zeta _3 \zeta _4+378 \zeta _7\biggr) \epsilon
   + \biggl(5868 \zeta _3 \zeta _5
   \nonumber\\
   &-\frac{14805 \zeta _8}{2}+1512 \zeta _{2,6}\biggr)\epsilon ^2-\biggl(732 \zeta _3^3+270 \zeta _4 \zeta _5-6930 \zeta _3 \zeta _6+\frac{42458 \zeta _9}{3}\biggr) \epsilon ^3
   \nonumber\\
   &
   +\biggl(-3294 \zeta _3^2 \zeta _4+\frac{202569 \zeta _5^2}{7}+\frac{697887 \zeta _3 \zeta _7}{7}-\frac{895521 \zeta _{10}}{7}-\frac{58563 \zeta _{3,7}}{7}\biggr)\epsilon ^4
   \nonumber\\
   &
   - \biggl(223152 \zeta _3^2 \zeta _5-105454 \zeta _5 \zeta _6+214815 \zeta _4 \zeta _7-\frac{1404105 \zeta _3 \zeta _8}{2}+1002552 \zeta _2 \zeta _9
   \nonumber\\
   &-\frac{4736453 \zeta _{11}}{4}+33504 \zeta _3 \zeta _{2,6}+54192 \zeta _{2,1,8}\biggr)\epsilon ^5
   + \biggl(-\frac{22816 \zeta _3^4}{3}-611040 \zeta _3 \zeta _4 \zeta _5
   \nonumber\\
   &-1041984 \zeta _2 \zeta _5^2-\frac{553066}{3} \zeta _3^2 \zeta _6-2118272 \zeta _2 \zeta _3 \zeta _7+\frac{17993629 \zeta _5 \zeta _7}{3}
   \nonumber\\
   &+\frac{60130438 \zeta _3 \zeta _9}{9}-\frac{659340762619 \zeta _{12}}{99504}-59016 \zeta _4 \zeta _{2,6}+77184 \zeta _2 \zeta _{3,7}
   \nonumber\\
   &-\frac{3234680 \zeta _{3,9}}{9}-120064 \zeta _{2,1,1,8}\biggr)\epsilon ^6+O\left(\epsilon ^7\right)
\Biggr\}\,,
\end{align}
\begin{align}
&M_{5,1}=\frac{(1-2 \epsilon )^3}{1+3 \epsilon}
\Biggl\{
   -\frac{5 \zeta _5}{\epsilon }-\biggl(17 \zeta _3^2+\frac{25 \zeta _6}{2}\biggr)-\biggl(51 \zeta _3 \zeta _4+\frac{85 \zeta _7}{2}\biggr) \epsilon
   \nonumber\\
   &
   +\biggl(\frac{138159 \zeta _8}{8}-11650 \zeta _3 \zeta _5-4266 \zeta _{2,6}\biggr)\epsilon ^2 +\biggl(8121 \zeta _4 \zeta _5-\frac{4310 \zeta _3^3}{3}-7710 \zeta _3 \zeta _6
   \nonumber\\
   &+\frac{16232 \zeta _9}{3}\biggr) \epsilon ^3
   + \biggl(-6465 \zeta _3^2 \zeta _4-\frac{863970 \zeta _5^2}{7}-\frac{572995 \zeta _3 \zeta _7}{7}+\frac{1796310 \zeta _{10}}{7}
   \nonumber\\
   &+\frac{171531 \zeta _{3,7}}{7}\biggr)\epsilon ^4
   + \biggl(456170 \zeta _3^2 \zeta _5-312737 \zeta _5 \zeta _6+530472 \zeta _4 \zeta _7-\frac{4475049 \zeta _3 \zeta _8}{4}
   \nonumber\\
   &+580086 \zeta _2 \zeta _9-\frac{2373525 \zeta _{11}}{4}+189612 \zeta _3 \zeta _{2,6}+31356 \zeta _{2,1,8}\biggr)\epsilon ^5
   + \biggl(\frac{196901 \zeta _3^4}{3}
   \nonumber\\
   &+1041474 \zeta _3 \zeta _4 \zeta _5+\frac{12645720}{7} \zeta _2 \zeta _5^2+\frac{35990}{3} \zeta _3^2 \zeta _6+\frac{25707760}{7} \zeta _2 \zeta _3 \zeta _7
   \nonumber\\
   &-\frac{38973620 \zeta _5 \zeta _7}{3}-\frac{77902346 \zeta _3 \zeta _9}{9}+\frac{2383042633405 \zeta _{12}}{199008}+393504 \zeta _4 \zeta _{2,6}
   \nonumber\\
   &-\frac{936720}{7} \zeta _2 \zeta _{3,7}+\frac{6770464 \zeta _{3,9}}{9}+208160 \zeta _{2,1,1,8}\biggr)\epsilon ^6+O\left(\epsilon ^7\right)
\Biggr\}\,,
\end{align}
\begin{align}
&M_{6,1}=\frac{(1-2 \epsilon )^3}{(1+3 \epsilon )^3}
\Biggl\{
   -\frac{10 \zeta _5}{\epsilon }-\biggl(100 \zeta _5+10 \zeta _3^2+25 \zeta _6\biggr)-\biggl(210 \zeta _5+100 \zeta _3^2
   \nonumber\\
   &+250 \zeta _6+30 \zeta _3 \zeta _4-\frac{19 \zeta _7}{2}\biggr) \epsilon
   +\biggl(-66 \zeta _3^2-525 \zeta _6-300 \zeta _3 \zeta _4+1257 \zeta _7
   \nonumber\\
   &+1564 \zeta _3 \zeta _5-567 \zeta _8+162 \zeta _{2,6}\biggr)\epsilon ^2
   + \biggl(-198 \zeta _3 \zeta _4+\frac{11151 \zeta _7}{2}+19344 \zeta _3 \zeta _5
   \nonumber\\
   &+1043 \zeta _8+972 \zeta _{2,6}+\frac{3440 \zeta _3^3}{3}+1374 \zeta _4 \zeta _5+3150 \zeta _3 \zeta _6+\frac{21637 \zeta _9}{3}\biggr)\epsilon ^3
   \nonumber\\
   &
   + \biggl(43092 \zeta _3 \zeta _5+28455 \zeta _8-1782 \zeta _{2,6}+\frac{40792 \zeta _3^3}{3}+23184 \zeta _4 \zeta _5+44000 \zeta _3 \zeta _6
   \nonumber\\
   &+\frac{339727 \zeta _9}{3}+5160 \zeta _3^2 \zeta _4+\frac{287486 \zeta _5^2}{7}-\frac{286 \zeta _3 \zeta _7}{7}+\frac{316935 \zeta _{10}}{7}+\frac{6642 \zeta _{3,7}}{7}\biggr)\epsilon ^4
   \nonumber\\
   &
   + \biggl(17848 \zeta _3^3+75330 \zeta _4 \zeta _5+116970 \zeta _3 \zeta _6+\frac{1061648 \zeta _9}{3}+61188 \zeta _3^2 \zeta _4
   \nonumber\\
   &+\frac{807571 \zeta _5^2}{2}+\frac{313421 \zeta _3 \zeta _7}{2}+\frac{2509185 \zeta _{10}}{4}+\frac{17127 \zeta _{3,7}}{2}-196432 \zeta _3^2 \zeta _5
   \nonumber\\
   &+158884 \zeta _5 \zeta _6+141174 \zeta _4 \zeta _7+133604 \zeta _3 \zeta _8+682428 \zeta _2 \zeta _9-\frac{17560877 \zeta _{11}}{24}
   \nonumber\\
   &
   -78360 \zeta _3 \zeta _{2,6}+36888 \zeta _{2,1,8}\biggr)\epsilon ^5+ \biggl(80316 \zeta _3^2 \zeta _4+\frac{13986207 \zeta _5^2}{14}+\frac{4863417 \zeta _3 \zeta _7}{14}
   \nonumber\\
   &+\frac{55755309 \zeta _{10}}{28}+\frac{490491 \zeta _{3,7}}{14}-1439456 \zeta _3^2 \zeta _5+1489454 \zeta _5 \zeta _6+\frac{3397821 \zeta _4 \zeta _7}{2}
   \nonumber\\
   &+925922 \zeta _3 \zeta _8+7235868 \zeta _2 \zeta _9-\frac{330988949 \zeta _{11}}{48}-583584 \zeta _3 \zeta _{2,6}+391128 \zeta _{2,1,8}
   \nonumber\\
   &-\frac{720896 \zeta _3^4}{9}-777104 \zeta _3 \zeta _4 \zeta _5-\frac{3772008}{7} \zeta _2 \zeta _5^2-\frac{1498208}{9} \zeta _3^2 \zeta _6+\frac{279408}{7} \zeta _2 \zeta _{3,7}
   \nonumber\\
   &-\frac{23004592}{21} \zeta _2 \zeta _3 \zeta _7+\frac{32644190 \zeta _5 \zeta _7}{9}+\frac{42933380 \zeta _3 \zeta _9}{27}+\frac{530242449679 \zeta _{12}}{298512}
   \nonumber\\
   &-219440 \zeta _4 \zeta _{2,6}-\frac{2444290 \zeta _{3,9}}{27}-\frac{186272}{3} \zeta _{2,1,1,8}\biggr)\epsilon ^6
   +O\left(\epsilon ^7\right)
\Biggr\}\,,
\end{align}

\textbf{Integrals with c.l.=4}
\begin{align}
&M_{6,2}=\frac{(1-2 \epsilon )^3}{(1+3 \epsilon )^3 (1+4 \epsilon)}
\Biggl\{
   -\frac{10 \zeta _5}{\epsilon }-\biggl(60 \zeta _5+10 \zeta _3^2+25 \zeta _6+70 \zeta _7\biggr)
   -  \biggl(50 \zeta _5
   \nonumber\\
   &-780 \zeta _3^2+150 \zeta _6+30 \zeta _3 \zeta _4+\frac{1829 \zeta _7}{2}-2560 \zeta _3 \zeta _5+4655 \zeta _8-1080 \zeta _{2,6}\biggr)\epsilon
   \nonumber\\
   &
   +\biggl(120 \zeta _5+6046 \zeta _3^2-125 \zeta _6+2340 \zeta _3 \zeta _4+7575 \zeta _7+29776 \zeta _3 \zeta _5-53193 \zeta _8
   \nonumber\\
   &+12258 \zeta _{2,6}+\frac{4528 \zeta _3^3}{3}-2640 \zeta _4 \zeta _5+1000 \zeta _3 \zeta _6-\frac{58460 \zeta _9}{9}\biggr)\epsilon ^2+\biggl(10704 \zeta _3^2
   \nonumber\\
   &
   +300 \zeta _6+18138 \zeta _3 \zeta _4+\frac{156783 \zeta _7}{2}+214920 \zeta _3 \zeta _5-215670 \zeta _8+59940 \zeta _{2,6}
   \nonumber\\
   &+\frac{73244 \zeta _3^3}{3}-28884 \zeta _4 \zeta _5+13200 \zeta _3 \zeta _6-\frac{1510937 \zeta _9}{18}+6792 \zeta _3^2 \zeta _4+\frac{143960 \zeta _5^2}{7}
   \nonumber\\
   &+\frac{495583 \zeta _3 \zeta _7}{7}-\frac{1038970 \zeta _{10}}{7}-\frac{88260 \zeta _{3,7}}{7}\biggr)\epsilon ^3
   + \biggl(32112 \zeta _3 \zeta _4+146088 \zeta _7
   \nonumber\\
   &+908472 \zeta _3 \zeta _5-395297 \zeta _8+167994 \zeta _{2,6}+140784 \zeta _3^3-37260 \zeta _4 \zeta _5+233700 \zeta _3 \zeta _6
   \nonumber\\
   &+\frac{166790 \zeta _9}{3}+109866 \zeta _3^2 \zeta _4+\frac{8380551 \zeta _5^2}{28}+\frac{26770505 \zeta _3 \zeta _7}{28}-\frac{108421795 \zeta _{10}}{56}
   \nonumber\\
   &-\frac{4631493 \zeta _{3,7}}{28}+390194 \zeta _3^2 \zeta _5-\frac{2438995 \zeta _5 \zeta _6}{6}+488786 \zeta _4 \zeta _7-\frac{1567351 \zeta _3 \zeta _8}{2}
   \nonumber\\
   &+3792685 \zeta _2 \zeta _9-\frac{100223975 \zeta _{11}}{16}-11366 \zeta _3 \zeta _{2,6}+205010 \zeta _{2,1,8}\biggr)\epsilon ^4
   \nonumber\\
   &
   + \biggl(1388592 \zeta _3 \zeta _5-324240 \zeta _8+211248 \zeta _{2,6}+\frac{1085348 \zeta _3^3}{3}+354744 \zeta _4 \zeta _5
   \nonumber\\
   &+1400980 \zeta _3 \zeta _6+\frac{45002569 \zeta _9}{18}+633528 \zeta _3^2 \zeta _4+\frac{15929910 \zeta _5^2}{7}+\frac{43860237 \zeta _3 \zeta _7}{7}
   \nonumber\\
   &-\frac{54746115 \zeta _{10}}{7}-\frac{6015150 \zeta _{3,7}}{7}+5634364 \zeta _3^2 \zeta _5-\frac{16554412 \zeta _5 \zeta _6}{3}
   \nonumber\\
   &+\frac{28467331 \zeta _4 \zeta _7}{4}-11209448 \zeta _3 \zeta _8+53532932 \zeta _2 \zeta _9-\frac{2821800989 \zeta _{11}}{32}
   \nonumber\\
   &-143668 \zeta _3 \zeta _{2,6}+2893672 \zeta _{2,1,8}-\frac{696554 \zeta _3^4}{9}-4241552 \zeta _3 \zeta _4 \zeta _5-\frac{42064920}{7} \zeta _2 \zeta _5^2
   \nonumber\\
   &+\frac{7951810}{9} \zeta _3^2 \zeta _6-\frac{256544080}{21} \zeta _2 \zeta _3 \zeta _7+\frac{254930897 \zeta _5 \zeta _7}{9}+\frac{714442631 \zeta _3 \zeta _9}{27}
   \nonumber\\
   &-\frac{12873185340379 \zeta _{12}}{597024}-843884 \zeta _4 \zeta _{2,6}+\frac{3115920}{7} \zeta _2 \zeta _{3,7}-\frac{43891225 \zeta _{3,9}}{27}
   \nonumber\\
   &-\frac{2077280}{3} \zeta _{2,1,1,8}\biggr)\epsilon ^5+O\left(\epsilon ^6\right)
\Biggr\}\,,
\end{align}
\begin{align}
&M_{6,3}=
\frac{(1-2 \epsilon )^3}{(1+3 \epsilon ) (1+4 \epsilon )}
\Biggl\{
   -\frac{5 \zeta _5}{\epsilon }-\biggl(20 \zeta _5+41 \zeta _3^2+\frac{25 \zeta _6}{2}-\frac{161 \zeta _7}{2}\biggr)
   \nonumber\\
   &
   +  \biggl(-308 \zeta _3^2-50 \zeta _6-123 \zeta _3 \zeta _4+514 \zeta _7+4862 \zeta _3 \zeta _5-\frac{24451 \zeta _8}{4}+1566 \zeta _{2,6}\biggr)\epsilon
   \nonumber\\
   &
   +\biggl(-924 \zeta _3 \zeta _4-1500 \zeta _7+68636 \zeta _3 \zeta _5-\frac{744639 \zeta _8}{8}+23220 \zeta _{2,6}+\frac{1526 \zeta _3^3}{3}
   \nonumber\\
   &-2103 \zeta _4 \zeta _5+4325 \zeta _3 \zeta _6+\frac{111709 \zeta _9}{36}\biggr)\epsilon ^2
   +\biggl(235200 \zeta _3 \zeta _5-\frac{710311 \zeta _8}{2}
   \nonumber\\
   &+85536 \zeta _{2,6}+\frac{22048 \zeta _3^3}{3}-36366 \zeta _4 \zeta _5+55695 \zeta _3 \zeta _6+\frac{237103 \zeta _9}{12}+2289 \zeta _3^2 \zeta _4
   \nonumber\\
   &+\frac{1341143 \zeta _5^2}{56}+\frac{3816969 \zeta _3 \zeta _7}{56}-\frac{7815019 \zeta _{10}}{112}-\frac{500565 \zeta _{3,7}}{56}\biggr)\epsilon ^3
   + \biggl(\frac{61040 \zeta _3^3}{3}
   \nonumber\\
   &-160416 \zeta _4 \zeta _5+161860 \zeta _3 \zeta _6-\frac{460411 \zeta _9}{9}+33072 \zeta _3^2 \zeta _4+\frac{60035137 \zeta _3 \zeta _7}{56}
   \nonumber\\
   &+\frac{12859479 \zeta _5^2}{56}-\frac{134815227 \zeta _{10}}{112}-\frac{7724781 \zeta _{3,7}}{56}-453668 \zeta _3^2 \zeta _5+\frac{280574047 \zeta _{11}}{64}
   \nonumber\\
   &+\frac{1346777 \zeta _5 \zeta _6}{6}-\frac{4654793 \zeta _4 \zeta _7}{8}+1309878 \zeta _3 \zeta _8-2749211 \zeta _2 \zeta _9-87752 \zeta _3 \zeta _{2,6}
   \nonumber\\
   &-148606 \zeta _{2,1,8}\biggr)\epsilon ^4
   + \biggl(91560 \zeta _3^2 \zeta _4+\frac{5319415 \zeta _5^2}{14}+\frac{55472425 \zeta _3 \zeta _7}{14}-\frac{705626 \zeta _3^4}{9}
   \nonumber\\
   &-\frac{7239597 \zeta _{3,7}}{14}-5183674 \zeta _3^2 \zeta _5+\frac{3338505 \zeta _5 \zeta _6}{2}-1597650 \zeta _{2,1,8}-\frac{1752859}{18} \zeta _3^2 \zeta _6
   \nonumber\\
   &+\frac{62741559 \zeta _3 \zeta _8}{4}-29556525 \zeta _2 \zeta _9+\frac{2990096591 \zeta _{11}}{64}-\frac{29828681054659 \zeta _{12}}{2388096}
   \nonumber\\
   &-\frac{53503881 \zeta _4 \zeta _7}{8}-\frac{142150835 \zeta _{10}}{28}-1729457 \zeta _3 \zeta _4 \zeta _5-\frac{25832277}{7} \zeta _2 \zeta _5^2
   \nonumber\\
   &-\frac{157544998}{21} \zeta _2 \zeta _3 \zeta _7+\frac{311533051 \zeta _5 \zeta _7}{18}+\frac{1792012205 \zeta _3 \zeta _9}{108}-1085340 \zeta _3 \zeta _{2,6}
   \nonumber\\
   &-227636 \zeta _4 \zeta _{2,6}+\frac{1913502}{7} \zeta _2 \zeta _{3,7}-\frac{105698899 \zeta _{3,9}}{108}-\frac{1275668}{3} \zeta _{2,1,1,8}\biggr)\epsilon ^5
   \nonumber\\
   &+O\left(\epsilon ^6\right)
\Biggr\}\,.
\end{align}

One may notice that rational coefficients in front of $\zeta_{12}$ involve the very big
prime number 691 in the denominator. This is because the same number appears in the
numerator of $\zeta_{12}=691\pi^{12}/638512875$. If we use $\pi^{12}$, instead of
$\zeta_{12}$, this number will disappear, and the highest prime factor in the coefficients
denominators will be only 13.

Our results are in the full agreement with the results of Ref.~\cite{Baikov:2010hf}
up to terms considered in that paper. Partly, this is because we used some of the data
from Ref.~\cite{Baikov:2010hf} for the determination of the homogeneous parts of the
solution of DRR. Nevertheless, our calculation can be considered as a nontrivial check
of the results of Ref.~\cite{Baikov:2010hf} because of many constraints on the terms
of $\epsilon$-expansion provided by the DRA method and fulfilled by the results of
Ref.~\cite{Baikov:2010hf}.

We also confirm the numerical results of Ref.~\cite{Smirnov:2010hd}
where one more term (as compared with Ref.~\cite{Baikov:2010hf}) of the
$\epsilon$-expansions of the most complicated thirteen master integrals was obtained
using {\tt FIESTA} \cite{FIESTA,FIESTA2}.

We observe that only MZV are present in our results for the four-loop
massless propagator master integrals. Since any other
four-loop massless propagator integral, with any
integer powers of numerators and propagators, can be represented, due to an IBP reduction,
as a linear combination of the master integrals, with coefficients which are rational
functions of $d$, we come to the conclusion that any four-loop massless propagator integral
has only MZV in its epsilon expansion up to transcendentality weight twelve. This means
that if we want to find something beyond MZV in four-loop
massless propagator diagrams we have to go to higher transcendentality weights. This is
certainly possible within our approach. In fact, we have chosen weight twelve because
it looks to be already a sufficiently big number. We can perform calculation up to higher
weights and will do our best on demand, for example, if somebody has reasons to
believe that unusual transcendental constants can appear at some specific weight.

This possibility is due to an important feature of the method that we use. Once we have a result in terms of a multiple series which always appears to be well convergent, going to higher powers of $\epsilon$ is an easy procedure, in contrast to other
methods.\footnote{For example, one of us (V.S.) remembers very well how painful it was to
obtain higher powers of $\epsilon$ when evaluating master integrals for three-loop form
factors by the method of MB representation~\cite{3lff3,3lff4}. In fact, two highest
coefficients were presented only in a numerical form at this time. Only one year later, they were evaluated analytically by the present method~\cite{Lee:2010cga}.}

Taking our results into account one obtains more motivations to try to prove that there are only MZV in massless propagator diagrams. Another alternative is to continue to look for unusual constants in higher loops. Keeping in mind the dramatic progress of the last years in the field of evaluating Feynman integrals, this also looks to be a possible scenario. All results presented here are available in computer-readable form on \\ %\href{http://www-ttp.particle.uni-karlsruhe.de/Progdata/ttp11/ttp11-20/}{www-ttp.particle.uni-karlsruhe.de/Progdata/ttp11/ttp11-20/}
\texttt{\small www-ttp.particle.uni-karlsruhe.de/Progdata/ttp11/ttp11-20/}.

\vspace{0.2 cm}

{\em Acknowledgments.}

We are grateful to P.A.~Baikov, K.G.~Chetyrkin and M.~Steinhauser for fruitful discussions
and careful reading of our paper.
This work was supported by the Russian Foundation for Basic Research through grant
11-02-01196 and by DFG through SFB/TR~9 ``Computational Particle Physics''.
The work of R.L. was also supported through Federal special-purpose program
``Scientific and scientific-pedagogical personnel of innovative Russia''.
R.L. gratefully acknowledges Karlsruhe Institut f\"{u}r Theoretische Teilchenphysik
for warm hospitality and financial support during his visit.

\end{document}